\documentclass{IEEEtran}
\usepackage{graphicx} 
\usepackage{authblk}
\usepackage{blindtext}
\usepackage{multicol}
\usepackage{cite}
\usepackage{amsmath,amssymb,amsfonts}
\usepackage{algorithmic}
\usepackage{textcomp}
\usepackage{xcolor}
\usepackage{xspace}
\usepackage{listings}

\begin{document}

\title{Transpiling RTL Pseudo-code of the POWER Instruction Set Architecture to C for Real-time Performance Analysis on Cavatools Simulator}
\author[1]{Kinar S}
\author[1]{Prashanth K V}
\author[1]{Adithya Hegde}
\author[1]{Aditya Subrahmanya Bhat}
\author[1]{Narender M}
\affil[1]{The National Institute of Engineering, Mysuru}

\maketitle

\begin{abstract}    
This paper presents a transpiler framework for converting RTL pseudo code of the POWER Instruction Set Architecture (ISA) to C code, enabling its execution on the Cavatools simulator. The transpiler consists of a lexer and parser, which parse the RTL pseudo code and generate corresponding C code representations. The lexer tokenizes the input code, while the parser applies grammar rules to build an abstract syntax tree (AST). The transpiler ensures compatibility with the Cavatools simulator by generating C code that adheres to its requirements. The resulting C code can be executed on the Cavatools simulator, allowing developers to analyze the instruction-level performance of the Power ISA in real time. The proposed framework facilitates the seamless integration of RTL pseudo code into the Cavatools ecosystem, enabling comprehensive performance analysis and optimization of Power ISA-based code.
\end{abstract}

\section{Introduction}
Efficient simulation and analysis of computer architectures is crucial for the design of high-performance systems. However, the disparity between desired system speed and the slow simulation means available poses a significant challenge. This discrepancy becomes especially pronounced when simulating multiprocessor systems, where the simulation rates lag far behind the capabilities of each core operating in the gigahertz range.~\cite{esaca}

To address this performance gap, this paper presents a novel transpiler framework that converts Power ISA RTL (Register Transfer Level) code to C code. This transpiler enables seamless integration of Power ISA RTL into the Cavatools ecosystem, leveraging its unique features to achieve high-performance simulation and analysis.

Cavatools is a simulator specifically designed for analyzing the performance of processors based on the POWER Instruction Set Architecture (ISA). It is concentrated on delivering precise and effective POWER ISA-based code execution, enabling in-depth instruction-level performance analysis. Cavatools is particularly suited for evaluating the behavior and optimization of Power ISA instructions in real-time scenarios. It offers a remarkable performance improvement of around 100X over existing Python-based simulators. This is achieved by breaking down the modeling of a single processor core into a pipeline of operations that can be simulated in parallel on a host computer with multiple cores. Additionally, Cavatools utilizes the RISC-V Weak Memory Ordering model to simulate multiple cores as parallel processes, synchronizing only when necessary for memory operations with defined system-wide ordering.~\cite{cava,simgem}

While PySim offers a versatile and customizable simulation platform for general computer systems research ~\cite{pysim}, Cavatools specializes in analyzing the performance of processors based on the POWER ISA.
\section{Background}

Also known as source-to-source compilers, \textbf{transpilers} are software tools that convert source code written in one programming language into an equivalent representation in another language while preserving the semantics and functionality of the original code. Transpilers serve as intermediaries for code translation, eliminating the need for manual rewriting. They enable code reuse, optimization, and platform adaptation.~\cite{transpiler}

Code simulators are essential tools in software development and computer science. They enable developers to create virtual environments that replicate the behavior and functionality of real systems. By running code within these simulated environments, developers can test and debug their software without the need for physical hardware. Simulators provide a safe and controlled space to observe and analyze code execution, identify potential issues, and optimize performance. They allow for rapid prototyping, enabling developers to experiment with different configurations and scenarios. Code simulators greatly enhance the development process by reducing costs, increasing productivity, and facilitating thorough testing and validation of software before deployment in real-world environments.

When it comes to simulators, C offers several advantages over Python that contribute to increased power and performance.

\begin{itemize}
    \item \textbf{Execution Speed}: C is a compiled language, whereas Python is an interpreted language. This fundamental difference in their execution models affects the speed of code execution. C code is compiled directly into machine code, resulting in highly optimized and efficient execution. On the other hand, Python code is interpreted at runtime, which introduces an additional layer of interpretation overhead. Consequently, C-based simulators tend to exhibit significantly faster execution speeds compared to Python-based simulators.
    \item \textbf{Low-Level Control}: C provides developers with direct access to low-level system resources and hardware, allowing for fine-grained control over memory management, data structures, and hardware-specific optimizations. This level of control is particularly advantageous for simulators that require precise timing, low latency, or real-time interaction with external devices. Python, being a higher-level language, abstracts away many low-level details, which can limit the level of control and optimization achievable in the simulation.
    \item \textbf{Memory Efficiency}: C offers explicit memory management capabilities through features such as manual memory allocation and de-allocation. This allows developers to optimize memory usage based on specific requirements, leading to efficient memory utilization and reduced memory overhead. Python, in contrast, utilizes automatic memory management through garbage collection, which introduces some overhead and may result in less efficient memory usage. In memory-intensive simulations, the ability to finely control memory allocation and de-allocation can have a significant impact on overall performance.
    \item \textbf{Cycle-Accurate Simulation}: Lower language run-time overhead and high optimizability allows for nearly cycle-accurate code simulation on C, while Python is bogged down by its own heavy runtime components, consuming more cycles for the same tasks.

\end{itemize}

\section{Design}
\subsection{Compiler Front-end}
The compiler's front-end is responsible for parsing and analyzing the RTL Power ISA code. It performs lexical analysis to tokenize the input code and then constructs an abstract syntax tree (AST) representing the structure of the code. The front-end verifies the syntax and semantics of the RTL Power ISA instructions, ensuring they conform to the ISA specification.
The parser is a fundamental component of a compiler for any Garden Snake-like indented language~\cite{gardensnake}. It facilitates the conversion of RTL source-code into code in C language. The parser follows a recursive descent parsing approach and relies on a lexer for tokenization and an emitter for code generation. It supports a range of statements, including assignments, if statements, switch statements, and do-while loops. Additionally, it handles the indentation-based block structures commonly encountered in Python-like languages.
\begin{lstlisting}[caption={Fixed Point Store Instruction RTL code.}, label={lst:fixedarith_RTL}]
if RA = 0 then
    b <-0
else
    b <- (RA)
EA <- b + EXTS(D)
\end{lstlisting}
The parser is capable of efficiently parsing expressions, encompassing arithmetic and comparison operations, and emits the corresponding code. It also maintains a set of symbols to ensure proper declaration and utilization of variables. By effectively analyzing and transforming RTL code, the parser serves as a crucial component~\cite{dragon} in the compilation process, enabling developers to seamlessly convert it to C.

\subsection{Code Generation}
Once the AST is constructed, the code generation phase translates the RTL Power ISA instructions into equivalent C code representations. Each RTL instruction is mapped to a sequence of C statements that mimic the behavior of the original RTL instruction. The code generation phase also handles data type conversions, function calls, and stack management as necessary.

\subsection{RISC Architecture Compatibility}
To ensure compatibility with the target RISC architecture, the generated C code needs to be adapted accordingly. This step involves mapping the RTL Power ISA's instruction formats, addressing modes, and memory models to their equivalents in the target RISC architecture. Additionally, any differences in the register set, calling conventions, and memory layout are addressed during this stage. It is essential to generate C code that can be compiled and executed correctly on the target architecture. The C code generated is compiled using the riscv64-unknown-linux-gnu-gcc compiler with the -static option as CAVATools simulates static files only.

\subsection{Integration with CAVATools Simulator}
Cavatools is a comprehensive simulator for multi-core RISC-V machines. The simulator is designed to run on an X86 Linux host and provides a user-mode multi-threaded Linux interface to the guest program. It leverages the standard RISC-V toolchain, such as GNU or LLVM, with GLIBC for compiling the guest program. Notably, the simulator supports all host Linux system calls, enabling seamless interaction between the guest program and the host environment.
The CAVATools simulator, which models a multi-core chip with single-issue in-order pipelines and private instruction and data level-1 caches, is used for real-time performance analysis.
The Cavatools suite consists of the following components:

\begin{enumerate}
  \item \textbf{uspike}: This component serves as a RISC-V instruction set interpreter. It utilizes Python scripts to extract instruction bit encoding and execution semantics from the official GitHub repository, ensuring accurate interpretation of RISC-V instructions.
  
  \item \textbf{caveat}: The performance simulator, caveat, models a multi-core chip with single-issue in-order pipelines and private instruction and data level-1 caches. It provides a detailed simulation of the performance characteristics of the multi-core RISC-V machine.
  
  \item \textbf{erised}: erised is a real-time performance viewer specifically designed for use with caveat. While the guest program is running, erised displays various performance metrics for each static instruction. These metrics include execution frequency, cycles per instruction, as well as instruction and cache misses. erised provides valuable insights into the program's performance during run-time.
\end{enumerate}

By combining these components, Cavatools offers a powerful and versatile simulation environment for multi-core RISC-V systems. It enables developers to analyze the performance of their guest programs, including multi-threaded programs compiled with Pthead and OpenMP, while benefiting from the availability of the complete Linux system call interface.
Building CAVATools involves building RISCV-tools which are RISCV-GNU-Toolchain, RISCV-PK (riscv proxy kernel), RISCV-ISA-SIM,
and RISCV-Opcodes. 

The transpiled C code is compiled for the target RISC architecture and executed on the CAVATools simulator while the guest program is running.The generated C code is compatible with the requirements of CAVATools. CAVATools captures detailed performance metrics for each static instruction, including execution frequency, cycles per instruction, instruction misses, and cache misses. These metrics provide low-level insights into the behavior of the RTL ISA and aid in performance analysis.~\cite{esaca}

\subsection{Performance Analysis}
The performance metrics collected by CAVATools can be used to analyze the RTL ISA's behavior on a low level. By examining the execution frequency, cycles per instruction, and cache-related statistics, performance bottlenecks, inefficient code segments, and cache inefficiencies can be identified. This analysis guides optimization efforts and helps improve the overall performance of the RTL ISA on the target RISC architecture.

\begin{figure}
    \centering
    \includegraphics[width=0.4\textwidth]{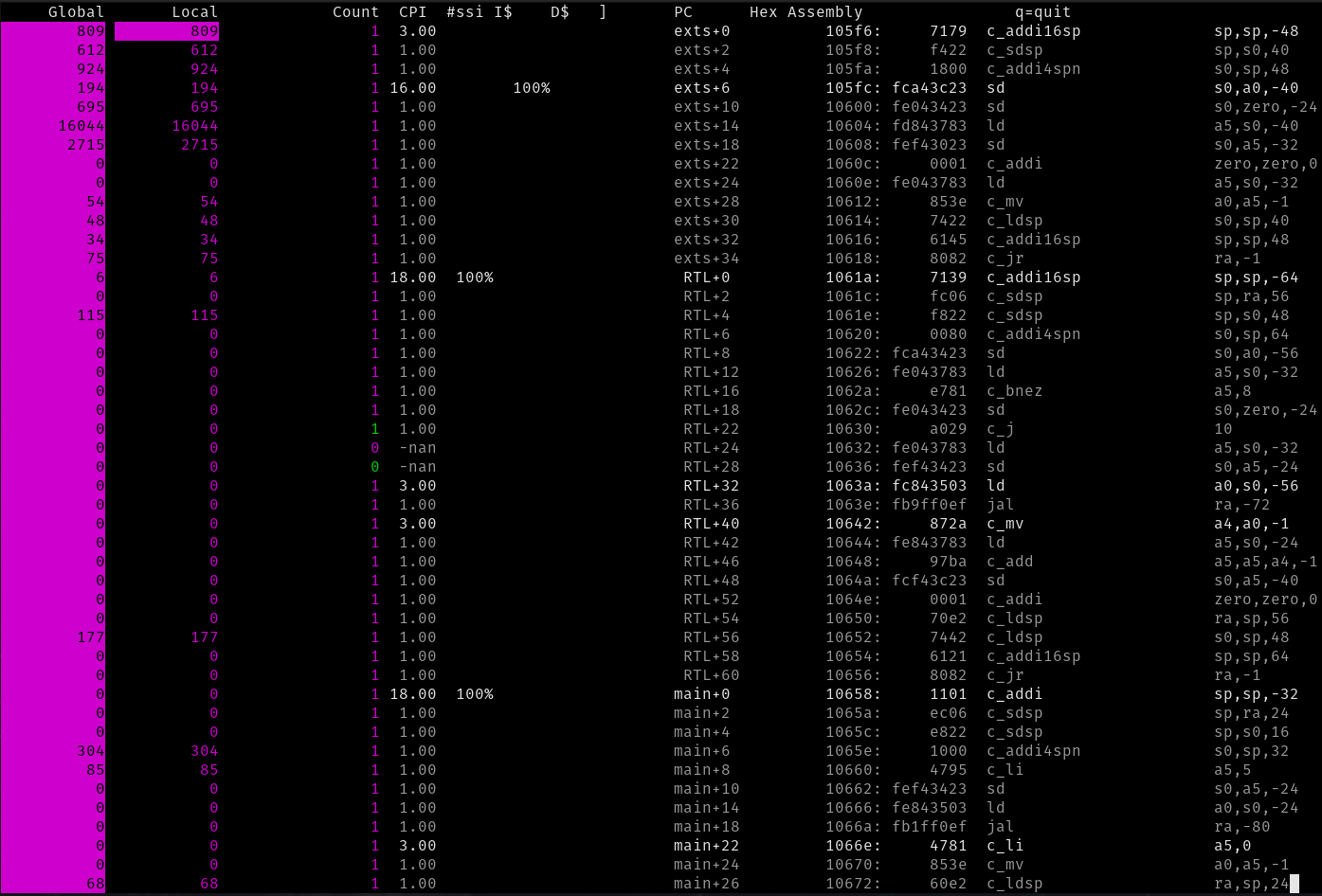}
    \caption{Erised displaying cache miss, CPI at instruction level in real time}
    \label{fig:erised output for fixed arithmetic RTL-C code}
\end{figure}

\section{Conclusion}
This paper introduces a transpiler framework that converts Power ISA RTL code to C code, enabling seamless integration with the Cavatools simulator. By combining the unique features of Cavatools with the transpiler framework, researchers and developers can harness the power of high-performance simulation and analysis for Power ISA-based systems, leading to enhanced system optimization and resource utilization.

By utilizing the transpiler framework, researchers and developers can seamlessly translate Power ISA RTL code into C code that is compatible with Cavatools. This approach enables efficient execution on the Cavatools simulator, capitalizing on its unique features and performance improvements. The transpiler preserves the integrity of the original Power ISA instructions, allowing for comprehensive performance analysis and optimization within the Cavatools environment.

The presented transpiler framework bridges the gap between Power ISA RTL code and the high-performance simulation capabilities of Cavatools. Through the transpilation to C code, developers gain access to the powerful simulation features of Cavatools, facilitating real-time performance analysis, architectural exploration, and system-level design.
\section*{Acknowledgement}
We would like to express our heartfelt gratitude to the organizations OpenPOWER and LibreSOC for their invaluable support. The collaborative efforts between our team and OpenPOWER and LibreSOC have allowed us to leverage their expertise, resources, and guidance, enabling us to tackle the challenges associated with translating RTL pseudo code to C code and integrating it seamlessly into the Cavatools ecosystem. The collaboration has not only enriched our research but also provided us with a deeper understanding of the intricacies of the POWER ISA.

Additionally, we would like to acknowledge the significant contributions made by the developer(s) behind Cavatools. The development of Cavatools has provided researchers and developers with a powerful simulation platform for analyzing instruction-level performance. The compatibility of our transpiler framework with Cavatools has allowed us to execute the generated C code and analyze the performance of Power ISA-based instructions in real-time.

We are deeply grateful for the opportunities, knowledge sharing, and technical support provided by OpenPOWER, LibreSOC, and the developer(s) behind Cavatools throughout the entire duration of this project. Their commitment to open-source, open collaboration, and technological advancement has been truly inspiring and has significantly contributed to the success of our work.
\bibliographystyle{IEEEtran}
\bibliography{main}

\end{document}